\documentclass[12pt]{article}

\usepackage[reqno]{amsmath}
\usepackage{mathrsfs}
\usepackage{amssymb}

\usepackage{rotating} 

\usepackage{bbm}
\usepackage{epsfig}
\usepackage{array}
\usepackage{float}
\usepackage{color}
\usepackage{graphicx}

\usepackage{a4}
\usepackage{a4wide}
\usepackage{wasysym}

\def\gs{\mathrel{
   \rlap{\raise 0.511ex \hbox{$>$}}{\lower 0.511ex \hbox{$\sim$}}}}
\def\ls{\mathrel{
   \rlap{\raise 0.511ex \hbox{$<$}}{\lower 0.511ex \hbox{$\sim$}}}}

\newcommand{\be}{\begin{eqnarray}}
\newcommand{\ee}{\end{eqnarray}}

\newcommand{\Rep}[1]{\underline{\mbox{\textbf{#1}}}}
\newcommand{\MoreRep}[2]{\underline{\mbox{\textbf{#1}}} _{\mbox{\textbf{#2}}}}

\begin{document}

\setlength{\unitlength}{1mm}

\begin{titlepage}
\title{\vspace*{-2.0cm}
\bf\Large
Confronting Flavour Symmetries and extended Scalar Sectors with Lepton Flavour Violation Bounds
\\[5mm]\ }

\author{
Adisorn Adulpravitchai\thanks{email: \tt adisorn$.$adulpravitchai@mpi-hd.mpg.de}~~,~~
Manfred Lindner\thanks{email: \tt manfred$.$lindner@mpi-hd.mpg.de}~~,~~and~~
Alexander Merle\thanks{email: \tt alexander$.$merle@mpi-hd.mpg.de}
\\ \\
{\normalsize \it Max-Planck-Institut f\"ur Kernphysik,}\\
{\normalsize \it Postfach 10 39 80, 69029 Heidelberg, Germany}
}
\date{}
\maketitle
\thispagestyle{empty}

\begin{abstract}
\noindent
We discuss the tension between discrete flavour symmetries and extended scalar sectors arising from lepton flavour violation experiments. The key point is that extended scalar sectors will generically lead to flavour changing neutral currents, which are strongly constrained by experiments. Because of the large parameter space in the scalar sector such models will, however, usually have no big problems with existing and future bounds (even though the models might be constrained). This changes considerably once a flavour symmetry is imposed in addition: Because of the symmetry, additional relations between the different couplings arise and cancellations become impossible in certain cases. The experimental bounds will then constrain the model severely and can easily exclude it. We consider two examples which show how these considerations are realized. The same logic should apply to a much wider class of models.
\end{abstract}

\end{titlepage}

\section{\label{sec:intro} Introduction}

The Standard Model (SM) is a very successful theory. Apart from only describing the phenomena observed, it has even made predictions of new particles such as the $Z$-boson or the $t$-quark. There is only one missing piece, namely the well-known neutral scalar Higgs boson, that will hopefully be found at the LHC collider.

Apart from this success there are, however, some observations which indicate that the SM is incomplete, among them, e.g., the observation of Dark Matter~\cite{Bertone:2004pz}, the baryon asymmetry~\cite{Riotto:1998bt}, and the hierarchy problem~\cite{Martin:1997ns}. An extension of the SM is therefore necessary (this is usually called Physics beyond the SM (BSM)). The various possible extensions of the SM very often contain extended scalar sectors. Other extensions address the flavour problem and introduce new flavour symmetries to explain the apparent regularities of masses and mixings.

In this paper we study the difficulties arising when one tries to combine a model with an extended scalar sector with a discrete flavour symmetry. The key point is that there are actually quite strong constraints on models with extended scalar sectors. Since, however, the scalar sector of a theory is in most of the cases poorly known (meaning that there are a lot of free parameters), such a model can usually not be excluded easily, because of internal cancellations between several of the parameters that may cause some observables to nearly vanish. If, on the other hand, some additional structure is imposed on the model (by, e.g., a discrete flavour symmetry), then additional relations between some of the parameters can easily rule out the corresponding model or to at least restrict its parameters to very narrow ranges.

The paper is organized as follows: In Sec.~\ref{sec:argumentation}, we introduce the argumentation which lead us to our statement that models with extended scalar sectors may get into trouble by the introduction of an additional flavour symmetry. This is exemplified in Sec.~\ref{sec:example}, where we present two particular models for which our logic clearly works. The numerical results that we have obtained are presented and discussed in Sec.~\ref{sec:results}, and we finally conclude in Sec.~\ref{sec:conclusions}. The basic properties of the discrete groups that we use are given in the Appendices A~($A_4$) and B~($D_4$).

\section{\label{sec:argumentation} The general arguments}

A natural way to extend the SM is to add further scalar particles, which have not yet been discovered. These could, e.g., be additional $SU(2)$-singlets~\cite{Barger:2007im}, doublets (``Two Higgs doublet model'', THDM), or triplets~\cite{Gunion:1989we}. Depending on the model, it can then be the case that more than one Higgs field contribute to the masses of all particles or that certain Higgses only give masses to a particular choice of particles~\cite{Haber:2006ue}. These models will then, however, generically lead to flavour changing neutral currents (FCNCs)~\cite{Diaz:2000cm} and hence to lepton flavour violation (LFV) processes~\cite{Blum:2007he}, which are quite strongly constrained~\cite{Raidal:2008jk}. It is, however, also not easy to rule them out that way, since they will in general yield complex $3\times 3$ Yukawa coupling matrices, which hold a lot of freedom in their 18 parameters. So, in most of the cases, such a model will be able to fit all neutrino data without any problems, even if it is strongly constrained.

On the other hand, there are also ways to impose more structure onto the SM in order to get an understanding of quantities like mixing angles, or so. This is usually done by so-called (discrete) ``flavour symmetries'' under which the SM-fermions (and, depending on the model, also (additional) scalars) are charged in a certain way. If, e.g., two generations of $SU(2)_L$ doublets are components of the same doublet representation of a discrete flavour symmetry (such as the dihedral groups $D_3 \simeq S_3$~\cite{Grimus:2005mu} or $D_4$ \cite{Grimus:2003kq,Ishimori:2008gp,D4paper}), then this property will generically lead to $\mu$-$\tau$ symmetry~\cite{Fukuyama:1997ky} by which two mixing angles are predicted, $\theta_{23}=\pi/4$ and $\theta_{13}=0$. Moreover, in order to increase the predictivity, one can also assign all three generations of $SU(2)_L$ doublets to form a triplet of a discrete flavour symmetry (such as $A_4$ \cite{Ma:2001dn,Babu:2002dz,Altarelli:2005yp,GAFF,Altarelli:2006kg}). This can lead to tri-bimaximal mixing~\cite{Harrison:2002er}, in which also $\theta_{12}$ is fixed to be $\tan \theta_{12}=1/\sqrt{2}$.

Imposing such symmetries adds more structure to the model in the sense that one obtains relations between different entries of the Yukawa matrices. By that way, one can obtain the neutrino oscillation parameters as well as the charged lepton masses as functions of only a few parameters, which can then be checked on whether they are in accordance with data, or not. However, such models generically need a lot of scalars in order to break the flavour symmetry in a valid way. In case the normal Higgses are not charged under the flavour symmetry, these are additional SM-singlet scalars (``flavons''), which are only charged under the discrete symmetry and can hence break it by obtaining a vacuum expectation value (VEV). These scalars will, again, lead to horribly large FCNCs, which crashes with phenomenology.

One way out is to decouple the flavons by giving them masses associated with the breaking scale of the flavour symmetry, which can be much higher then the electroweak scale. This is, of course, somehow only hiding the problem, but it will make the model fit better.\\

We now apply the following logic:
\begin{enumerate}

\item We impose a flavour symmetry and decouple the flavons in order to end up with an effective low energy model with a scalar sector that is slightly extended compared to the SM. This could, e.g., be a THDM or something similar.

\item This procedure should make the model fit better, because the possible problems that could arise by the flavons are avoided.

\item Since we have gained predictivity by imposing the flavour symmetry, we can fit the model to neutrino data, which allows us to extract certain ranges for the model parameters.

\item The model, however, still has additional scalars compared to the SM, which will be able to mediate LFV-processes, whose branching ratios can be predicted using the fitted parameter values.

\item If this prediction does not fit with present (future) LFV-bounds, we are (will be) able to exclude the particular flavour symmetry imposed (in a certain scenario). Note that this logic will also hold in the non-decoupling case if no extreme fine-tuning is involved.

\end{enumerate}

In principle, this could work for any model with a slightly extended scalar sector. If the structure of the model is not extremely peculiar, which is rarely the case in the scalar sector of a theory, the additional scalars (compared to the SM) will unavoidably lead to LFV-processes, which are already strongly constrained. The key point is that these constraints are so strong, that imposing some more structure by adding a flavour symmetry can easily destroy the consistency of the model with all data.

Here, we want to present such an analysis for one particular example, namely for Ma's scotogenic model~\cite{Ma:2006km}, as this consists of a very minimal extension of the SM. Furthermore, it does not have too many possible LFV-diagrams, so that our logic is not shadowed by a heavy calculational apparatus. In this model, one can see immediately the effect of certain symmetries: Without imposing a flavour symmetry, one constrains quantities like
\begin{equation}
 |h_{11}^* h_{21} + h_{12}^* h_{22} + h_{13}^* h_{23}|
 \label{eq:meg-constraint}
\end{equation}
by LFV-processes like $\mu \rightarrow e\gamma$~\cite{Sierra:2008wj}, where $h$ is the Yukawa coupling matrix involved. Such a combination can easily become zero for unfortunate values of some phases, exactly as the effective neutrino mass in neutrino-less double beta processes~\cite{Lindner:2005kr}. Imposing relations between certain elements of $h$ hinders such cancellations to appear, and the term in Eq.~\eqref{eq:meg-constraint} will generically be much larger than zero.

We want to stress, however, that this particular model is just an example and that our idea may work for a much wider class of models.

\section{\label{sec:example} Constraining particular models}

\subsection{\label{sec:Ma-model} One possible example: The scotogenic model}

There are a lot of different models for neutrino mass generation on the market~\cite{Ma:2009dk}. A difficult task for all of them is to explain the smallness of neutrino masses compared to other particles we know in Nature.

One way is to forbid a tree-level mass term for neutrinos and generate neutrino masses only by radiative corrections, as done in several models \cite{Ma:2006km,Ma:2009dk,Zee:1980ai,Babu:1988ig,Zee:1985id,Babu:1988ki}. Out of those, Ma's ``scotogenic'' model~\cite{Ma:2006km} (that we call ``Ma-model'' for simplicity) is particularly attractive: By adding only one additional Higgs doublet and heavy right-handed neutrinos to the SM, as well as imposing an additional $Z_2$-symmetry, it allows for sufficiently small neutrino masses. These masses are generated radiatively, because the additional neutral Higgs does not obtain a VEV that could lead to a tree-level neutrino mass term. Furthermore, due to the $Z_2$-symmetry, this model also provides a stable Dark Matter candidate, namely the lightest of the heavy neutrinos~\cite{Suematsu:2009ww} or the lightest neutral scalar~\cite{Dolle:2009fn}. Constraints on the model arise from various different sources as, e.g., lepton flavour violation or the Dark Matter abundance~\cite{Sierra:2008wj}. In that sense, this model is very ``complete''.

The basic ingredients apart from the SM are:
\begin{itemize}

\item 3 heavy right-handed (Majorana) neutrinos $N_k$, which are singlets under $SU(2)$ and have no hypercharge

\item a second Higgs doublet $\eta$ with SM-like quantum numbers that does not obtain a VEV

\item an additional $Z_2$-parity under which all SM-particles are even, while $N_k$ as well as $\eta$ are odd

\end{itemize}

The corresponding Higgs potential looks like
\begin{equation}
 V= m_1^2 \phi^\dagger \phi + m_2^2 \eta^\dagger \eta + \frac{\lambda_1}{2} (\phi^\dagger \phi)^2+ \frac{\lambda_2}{2} (\eta^\dagger \eta)^2+ \lambda_3 (\phi^\dagger \phi)(\eta^\dagger \eta)+ \lambda_4 (\phi^\dagger \eta) (\eta^\dagger \phi)+ \frac{\lambda_5}{2} \left[ (\phi^\dagger \eta)^2 +h.c.\right],
 \label{eq:Higgs_pot}
\end{equation}
where $\phi$ is the SM-Higgs. If $m_1^2<0$ and $m_2^2>0$, then only $\phi^0$ will obtain a VEV $v=174$~GeV, while $\langle \eta^0 \rangle=0$. Then, the Yukawa Lagrangian is given by
\begin{equation}
 \mathcal{L}_Y=f_{ij} (\phi^- \nu_i + (\phi^0)^* l_i) e_j^c+ h_{ij} (\eta^0 \nu_i - \eta^+ l_i) N_j+h.c.,
 \label{eq:Yukawa}
\end{equation}
which does not lead to a tree-level neutrino mass term due to the vanishing VEV of $\eta^0$. The neutrino masses can, however, be generated radiatively, which gives a natural suppression of the neutrino mass eigenvalues and can exploit the heaviness of the $N_k$ (with masses $M_k$) as well. The mass matrix of the light neutrinos reads
\begin{equation}
 (\mathcal{M}_\nu)_{ij}=\sum_{k=1}^3 h_{ik} h_{jk} \Lambda_k,
 \label{eq:nu-mass-matrix}
\end{equation}
where
\begin{equation}
 \Lambda_k=\frac{M_k}{16 \pi^2} \left[ \frac{m^2(H^0)}{m^2(H^0)-M_K^2} \ln \left( \frac{m^2(H^0)}{M_K^2} \right)- \frac{m^2(A^0)}{m^2(A^0)-M_K^2} \ln \left( \frac{m^2(A^0)}{M_K^2} \right) \right].
 \label{eq:Lambdak}
\end{equation}
Note that we have named the Higgses like in the general THDM, with $\alpha=\beta=m_{12}=\lambda_{6,7}=0$~\cite{Eriksson:2009ws}. The resulting Higgs masses are given by
\begin{eqnarray}
 &&m^2(h^0)=2\lambda_1 v^2,\ m^2(H^0)=m_2^2+(\lambda_3+\lambda_4+\lambda_5)v^2,\ m^2(A^0)=m_2^2+(\lambda_3+\lambda_4-\lambda_5)v^2,\nonumber\\
 &&{\rm and}\ m^2(H^\pm)=m_2^2+\lambda_3 v^2.
 \label{eq:Higgses}
\end{eqnarray}

\subsection{\label{sec:models} The flavour symmetries considered}

In the following, we will present two models which constrain the structure of the Yukawa coupling matrix $h$ in Eq.~\eqref{eq:Yukawa}, without discussing a particular mechanism for vacuum alignment.~\footnote{In general, the vacuum alignment can be achieved by a minimization of the scalar potential.} The first one, based on Ref.~\cite{A4seesawMuandKing}, represents the class of models which predicts tri-bimaximal mixing. The second one represents the class which predicts $\mu-\tau$ symmetry.

\subsubsection{\label{sec:A4} The $A_4$-model (Model 1)}
\begin{table}
\begin{center}
\begin{tabular}{|c||c|c|c|c|c||c|c||c|c|c|}\hline
Field & $l_{1,2,3}$ & $e^{c}_{1}$ &
$e^{c}_{2}$ & $e^{c}_{3}$ 
& $N_{1,2,3}$ & $\phi$ & $\eta$ & $\varphi_S$ & $\varphi_T$ & $\chi$ \\
\hline
$A_4$ & $\Rep{3}$ &
$\Rep{1}$ & $\Rep{1}''$ & $\Rep{1}'$ & $\Rep{3}$ & $\Rep{1}$
 & $\Rep{1}$ & $\Rep{3}$ & $\Rep{3}$ & $\Rep{1}$ \\
$Z_4^{\rm aux}$ & $i$ & $i$ & 
$i$ & $i$ & $-1$ & $1$ & $1$ & $i$ & $-1$
& $i$ \\
\hline
\end{tabular}
\end{center}
\begin{center}
\begin{minipage}[t]{12cm}
\caption{\label{tab:particles1} The particle content of Model 1: The SM particles are the three left-handed lepton $SU(2)_L$ doublets $l_i$, the right-handed charged leptons $e^c_i$, and the SM-Higgs $\phi$. The BSM particles are the right-handed neutrinos $N_i$, the second Higgs doublet $\eta$ (which does not obtain a VEV), and the flavons $\varphi_S$, $\varphi_T$, and $\chi$, that only transform under $A_4 \times Z_4^{\rm aux}$.
}
\end{minipage}
\end{center}
\end{table}

The particle content of this model is given in Tab.~\ref{tab:particles1}. The Lagrangian which is invariant under the flavour symmetry $A_4 \times Z_4^{\rm aux}$ reads\footnote{Here, we neglect the anti-symmetric part of the coupling between $l$ and $N$ or assume that the anti-symmetric coupling vanishes, which is done similarly in Refs.~\cite{A4seesawMuandKing,Hirsch:2008rp}.}
\begin{eqnarray}
\mathcal{L}_l &=& y_1^e \frac{\phi}{\Lambda} (l_1 \varphi_{T1} +l_2 \varphi_{T3}+l_3 \varphi_{T2}) e^c_1 + y_2^e \frac{\phi}{\Lambda} (l_3 \varphi_{T3} +l_1 \varphi_{T2}+l_2 \varphi_{T1}) e^c_2 \nonumber \\
& &+ y_3^e \frac{\phi}{\Lambda} (l_2 \varphi_{T2} +l_1 \varphi_{T3}+l_3 \varphi_{T1}) e^c_1 + \frac{\eta}{\Lambda} \Big[ y_1[(2 l_1 N_1 -l_2 N_3-l_3 N_2) \varphi_{S1} \nonumber \\
& & + (2 l_3 N_3 -l_1 N_2-l_2 N_1) \varphi_{S3} + (2 l_2 N_2 -l_1 N_3-l_3 N_1) \varphi_{S2}]  \nonumber \\
& &+ y_2 (l_1 N_1 + l_2 N_3 + l_3 N_2) \chi \Big]+M(N_1 N_1 + N_2 N_3 + N_3 N_2).
\end{eqnarray}
Let us assume that the flavons obtain their VEVs as follows,
\begin{equation}
\left( \begin{array}{c} \langle \varphi_{S1} \rangle \\ \langle \varphi_{S2} \rangle \\ \langle \varphi_{S3} \rangle \end{array} \right) 
=  w_S \, \left( \begin{array}{c} 1 \\ 1 \\1 \end{array} \right) \; , \;\;
\left( \begin{array}{c} \langle \varphi_{T1} \rangle \\ \langle \varphi_{T2} \rangle \\ \langle \varphi_{T3} \rangle \end{array} \right) 
=  w_T \, \left( \begin{array}{c} 1 \\ 0 \\0 \end{array} \right) \; , \;\;{\rm and}\;\;
\langle \chi \rangle = u \; , \;\;
\label{FlavonVEVmodel1}
\end{equation}
and the SM Higgs gets the VEV $\langle \phi \rangle = v$. Then, the Yukawa coupling matrix and the right-handed neutrino mass matrix for Model 1 can be written as
\begin{equation}
h=
 \begin{pmatrix}
 2a+b & -a & -a\\
 -a & 2a & b-a\\
 -a & b-a & 2a
 \end{pmatrix}
\; {\rm and}\;\;
M_{R}=M
 \begin{pmatrix}
 1 & 0 & 0\\
 0 & 0 & 1\\
 0 & 1 & 0
 \end{pmatrix},
 \label{eq:h_and_MR_model_1}
\end{equation}
where $a=y_1 \frac{w_S}{\Lambda}$ and $b=y_2 \frac{u}{\Lambda}.$ \\
The charged lepton mass matrix in this model is diagonal,
\begin{equation}
m_e= \frac{v}{\Lambda} y_1^e w_T , \; m_{\mu}= \frac{v}{\Lambda} y_2^e w_T, \; m_{\tau}= \frac{v}{\Lambda} y_3^e w_T.
\end{equation}
Here, the hierarchies in the charged lepton masses are determined by the Yukawa couplings. Assuming that the Yukawa coupling of the $\tau$, $y_3^e$, is of $\mathcal{O}(1)$ and the Higgs VEV $v$ is $174$~GeV, we can determine the ratio of the flavon over the cutoff scale $\Lambda$ $(\frac{\langle f \rangle}{\Lambda})$ as being of the order of the Cabibbo angle squared, $\lambda^2 \sim 0.04$.

In order to make the discussion easier, we go to the basis where the right-handed neutrino mass matrix is diagonal. The matrix $M_R M_R^{\dagger}$ is diagonalized by the unitary matrix $U_r$
\begin{equation}
U_{r}=
 \begin{pmatrix}
 0 & 0 & 1\\
 0 & 1 & 0\\
 1 & 0 & 0
 \end{pmatrix}.
\end{equation}
Note that the right-handed neutrino masses are degenerate, $M_{1,2,3}=M$. \\
The Yukawa coupling in this basis reads
\begin{equation}
h'=h U_r=
\begin{pmatrix}
 -a & -a & 2a+b\\
 b-a & 2a & -a\\
 2a & b-a & -a
 \end{pmatrix}.
\end{equation}
Using Eq.~\eqref{eq:nu-mass-matrix}, the neutrino mass matrix can be written as
\begin{equation}
M_{\nu}=\Lambda_{1,2,3}
 \begin{pmatrix}
 (6a^2+4ab+b^2) & -a(3a+2b) & -a(3a+2b)\\
 -a(3a+2b) & (6a^2-2ab+b^2) & a(-3a+4b)\\
 -a(3a+2b) & a(-3a+4b) & (6a^2-2ab+b^2)
 \end{pmatrix},
 \label{model_1 neutrino mass matrix}
\end{equation}
where $\Lambda_{1,2,3}=\Lambda_1=\Lambda_2=\Lambda_3$, and the neutrino masses are given by the eigenvalues of $M_{\nu} M_{\nu}^{\dagger}$:
\begin{equation}
m_1^2=(3a+b)^4\Lambda_{1,2,3}^2, \;\; m_2^2=b^4 \Lambda_{1,2,3}^2, \;\; \mbox{and} \; m_3^2=(-3a+b)^4 \Lambda_{1,2,3}^2,
\end{equation}
which correspond to the eigenvectors $(-2,1,1)^T/\sqrt{6}$, $(1,1,1)^T/\sqrt{3}$, and $(0,-1,1)^T/\sqrt{2}$, respectively. In this model, the neutrino masses obey normal mass ordering.

The neutrino mixing observables look like:
\begin{equation}
 \Delta m_\odot^2=(b^4-(3a+b)^4)\Lambda_{1,2,3}^2,\ \Delta m_A^2=-24ab(9a^2+b^2)\Lambda_{1,2,3}^2,\ \tan \theta_{12}=\frac{1}{\sqrt{2}},\ \theta_{13}=0,\ {\rm and}\ \theta_{23}=\frac{\pi}{4}.
 \label{eq:neutrino-observables_1}
\end{equation}
In this model, we have only three free parameters ($a,b,M$) to fit all observables. Therefore, this model is quite predictive (and hence harder to fit).

\subsubsection{\label{sec:D4} The $D_4$-model (Model 2)}
\begin{table}
\begin{center}
\begin{tabular}{|c||c|c|c|c|c|c|c||c|c||c|c|c|c|}\hline
Field & $l_{1}$ & $l_{2,3}$ & $e^{c}_{1}$ &
$e^{c}_{2,3}$ & $N_{1}$
& $N_{2}$ & $N_{3}$ & $\phi$ & $\eta$
& $\varphi_e$ & $\chi_e$ & $\varphi_{\nu}$ & $\psi_{1,2}$ \\
\hline
$D_4$ & $\MoreRep{1}{1}$ &
$\MoreRep{2}{}$ & $\MoreRep{1}{3}$ & $\MoreRep{2}{}$ & $\MoreRep{1}{3}$ & $\MoreRep{1}{2}$
 & $\MoreRep{1}{4}$ & $\MoreRep{1}{1}$ & $\MoreRep{1}{1}$ & $\MoreRep{1}{3}$ & $\MoreRep{1}{4}$& $\MoreRep{1}{3}$
& $\MoreRep{2}{}$ \\
$Z_2^{\rm aux}$ & $1$ & $1$ & 
$1$ & $1$ & $-1$ & $-1$ & $-1$ & $1$ & $1$
& $1$ & $1$ & $-1$ & $-1$\\
\hline
\end{tabular}
\end{center}
\begin{center}
\begin{minipage}[t]{12cm}
\caption{\label{tab:particles2} The particle content of Model 2: Th SM particles are the three left-handed lepton $SU(2)_L$ doublets $l_i$,
the right-handed charged leptons $e^c_i$, and the SM-Higgs $\phi$. The BSM particles are the right-handed neutrinos $N_i$, second Higgs doublet $\eta$ (which does not obtain a VEV),and the flavons $\varphi_e$, $\chi_e$, $\varphi_{\nu}$, and $\psi_i$, that only transform
under $D_4 \times Z_2^{\rm aux}$.}
\end{minipage}
\end{center}
\end{table}

The particle content of this model is given in Tab.~\ref{tab:particles2}. The Lagrangian which is invariant under the flavour symmetry $D_4 \times Z_2^{\rm aux}$ reads
\begin{eqnarray}
\mathcal{L}_l &=& y_1^e l_1 e_1^c \frac{\phi}{\Lambda} \varphi_e + y_2^e (l_2 e_2^c + l_3 e^c_3) \frac{\phi}{\Lambda} \varphi_e
       + y_3^e (l_2 e_2^c - l_3 e^c_3) \frac{\phi}{\Lambda} \chi_e \nonumber \\
    & & + y_1 l_1 N_1 \frac{\eta}{\Lambda} \varphi_{\nu} + y_2 (l_2 \psi_1 + l_3 \psi_2) N_1 \frac{\eta}{\Lambda} + y_3 (l_2 \psi_2 - l_3 \psi_1) N_2 \frac{\eta}{\Lambda}  +y_4 (l_2 \psi_1 - l_3 \psi_2) N_3 \frac{\eta}{\Lambda} \nonumber \\
    & & + \frac{1}{2} M_1 N_1 N_1 + \frac{1}{2} M_2 N_2 N_2 + \frac{1}{2} M_3 N_3 N_3.
\end{eqnarray}
Let us assume that the flavons obtain their VEVs as follows:
\begin{equation}
\langle \varphi_{e} \rangle = u_e, \;\; \langle \chi_{e} \rangle = -w_e , \;\;\langle \varphi_{\nu} \rangle = u \; , \;\;{\rm and} \;\;
\left( \begin{array}{c} \langle \psi_1 \rangle \\ \langle \psi_2 \rangle \end{array} \right)
=  w \, \left( \begin{array}{c} 1 \\ -1 \end{array} \right), \label{FlavonVEVmodel2}
\end{equation}
and the SM Higgs gets the VEV $\langle \phi \rangle = v$. Then, the Yukawa coupling matrix for Model 2 can be written as
\begin{equation}
h=
 \begin{pmatrix}
 a & 0 & 0\\
 b & -c & d\\
 -b & -c & d
 \end{pmatrix}
 \label{eq:h_model_2} ,
\end{equation}
where $a=y_1 \frac{u}{\Lambda}$, $b=y_2 \frac{w}{\Lambda}$, $c=y_3 \frac{w}{\Lambda}$, and $d=y_4 \frac{w}{\Lambda}.$

The charged lepton and right-handed neutrino mass matrices in this model are diagonal,
\begin{equation}
m_e= \frac{v}{\Lambda} y_1^e u_e , \; m_{\mu}= \frac{v}{\Lambda} (y_2^e u_e - y_3^e w_e), \; m_{\tau}= \frac{v}{\Lambda} (y_2^e u_e + y_3^e w_e).
\end{equation}
Here, the hierarchy between the masses of $e$ and $(\mu,\tau)$ arises from the smallness of the yukawa coupling $y_1^e$.
As we did for Model 1, we assume that the ratio $(\frac{\langle f \rangle}{\Lambda})$ is of order $\lambda^2 \sim 0.04$.
 
Using Eq.~\eqref{eq:nu-mass-matrix}, the neutrino mass matrix can be written as 
\begin{equation}
M_{\nu}=
 \begin{pmatrix}
 a^2 \Lambda_1 & a b \Lambda_1 & -a b \Lambda_1\\
 a b \Lambda_1 & b^2 \Lambda_1 + c^2 \Lambda_2 + d^2 \Lambda_3 & -b^2 \Lambda_1 + c^2 \Lambda_2 + d^2 \Lambda_3\\
 -a b \Lambda_1 & -b^2 \Lambda_1 + c^2 \Lambda_2 + d^2 \Lambda_3 & b^2 \Lambda_1 + c^2 \Lambda_2 + d^2 \Lambda_3
 \end{pmatrix}
 \label{model_2 neutrino mass matrix}.
\end{equation}
The neutrino masses are given by the eigenvalues of $M_{\nu} M_{\nu}^{\dagger}$,
\begin{equation}
m_1^2=0, \;\; m_2^2=(a^2+2 b^2)^2 \Lambda_1^2, \;\; \mbox{and} \; m_3^2=4 (c^2 \Lambda_2 + d^2 \Lambda_3)^2,
\end{equation}
which correspond to the eigenvectors
\begin{equation}
\frac{a}{\sqrt{2(a^2+2 b^2)}}(2b/a,-1,1)^T,\ \frac{b}{\sqrt{2(a^2+2 b^2)}}(-b/a,1,1)^T,\ {\rm and}\ (0,1,1)^T/\sqrt{2},
 \label{eq:EV_2}
\end{equation}
respectively.

In this model, the neutrino masses will obey normal ordering.

The neutrino mixing observables look like:
\begin{equation}
 \Delta m_\odot^2=(a^2+2b^2)^2\Lambda_1^2,\ \Delta m_A^2=4 (c^2 \Lambda_2 + d^2 \Lambda_3)^2,\ \tan \theta_{12}=\frac{a}{\sqrt{2} b},\ \theta_{13}=0,\ {\rm and}\ \theta_{23}=\frac{\pi}{4}.
 \label{eq:neutrino-observables_2}
\end{equation}
In this model, we have 7 free parameters ($a,b,c,d,M_1,M_2,M_3$) to fit all neutrino observables. This makes Model 2 much easier to fit, but we of course pay the price of losing predictivity.

\subsection{\label{sec:procedure} Phenomenological analysis}
\begin{table}[t]
\centering
\begin{tabular}[h]{|c||c|c|c|c|}\hline
Scenario & $m(h^0)$ & $m(H^0)$ & $m(A^0)$ & $m(H^\pm)$\\ \hline \hline
$\alpha$ & 120.0 &  32.9 &  84.5 &  93.0\\ \hline
$\beta$  & 120.0 &  60.4 & 101.5 & 111.5\\ \hline
$\gamma$ & 120.0 & 946.8 & 950.0 & 950.3\\ \hline
$\delta$ & 120.0 & 548.9 & 549.4 & 550.6\\ \hline
\end{tabular}
\caption{\label{tab:Higgs-masses} The Higgs masses (in GeV) for the different scenarios defined in Eq.~\eqref{eq:Higgs-scenarios}.}
\end{table}

\begin{table}[t]
\centering
\begin{tabular}[h]{|c||c|c|c|c|c|}\hline
Quantity & $\Delta m_\odot^2$ & $(\Delta m_A^2)_{\rm nor.}$ & $\theta_{12}$ & $\theta_{13}$ & $\theta_{23}$\\ \hline \hline
Best-fit   & $7.67\cdot 10^{-5}~{\rm eV}^2$ & $2.46\cdot 10^{-3}~{\rm eV}^2$ & $34.5^\circ$ & $0.0^\circ$ & $42.3^\circ$\\ \hline
$1\sigma$  & $2.15\cdot 10^{-6}~{\rm eV}^2$ & $0.15\cdot 10^{-3}~{\rm eV}^2$ & $1.4^\circ$ & $7.9^\circ$ & $4.2^\circ$\\ \hline
\end{tabular}
\caption{\label{tab:neutrino_params} The neutrino mixing parameters (best-fit values and symmetrized 1$\sigma$-ranges) obtained by a global fit~\cite{GonzalezGarcia:2007ib}.}
\end{table}

\subsubsection{\label{sec:general} The general procedure}

In this section, we describe the analysis procedure we have applied. The first thing to say is that there are constraints that are required for a THDM like in Eq.~\eqref{eq:Higgs_pot} ($\lambda_1>0$, $\lambda_2>0$, $\lambda_3>-\sqrt{\lambda_1 \lambda_2}$, and $\lambda_3+\lambda_4-|\lambda_5|>-\sqrt{\lambda_1 \lambda_2}$; they keep the potential stable) as well as consistency conditions for a Ma-like model ($m_1^2<0$ and $m_2^2>0$; these are necessary in order for $\phi^0$ to obtain a VEV, while $\eta^0$ obtains none). Furthermore, there are limits from direct searches at collider experiments~\cite{Raspereza:2002ni}: $m(h^0)>112.9$~GeV and $m(H^\pm)>78.6$~GeV, both at $95\%$ confidence level.\footnote{Note that these constraints do not apply to the ``inert'' Higgses $H^0$ and $A^0$. They are constrained much less severely by the current limits, differently from a normal THDM.} Further constraints arise from the $W$- and $Z$-boson decay widths, namely $m(H^\pm)+m(H^0), m(H^\pm)+m(A^0) > M_W$ and $2m(H^\pm), m(H^0)+m(A^0)>M_Z$, as well as from the requirement of perturbativity for the Higgs potential, $\lambda_2<1$ and $\lambda_3^2+(\lambda_3+\lambda_4)^2+\lambda_5^2<12\lambda_1^2$~\cite{Dolle:2009fn}.

Strong constraints also come from the correction to the $\rho$-parameter~\cite{Grimus:2007if}. The explicit formula for this correction reads
\begin{equation}
 \Delta \rho=\frac{\alpha(M_Z)}{16\pi s_W^2 M_W^2}\cdot \left[F(m_2^2,m^2(H^0))+F(m_2^2,m^2(A^0))-F(m^2(H^0),m^2(A^0)) \right],
 \label{eq:Delrho_Ma}
\end{equation}
where
\begin{equation}
 F(x,y)=\left\{
\begin{matrix}
\frac{x+y}{2}-\frac{xy}{x-y} \ln \frac{x}{y},\ {\rm for}\ x\neq y,\\
0,\ {\rm for}\ x= y,\hfill \hfill
\end{matrix}
 \right.
 \label{eq:F-function}
\end{equation}
and $\alpha(M_Z)=1/127.9$. The experimental constraint is~\cite{Amsler:2008zzb}
\begin{equation}
 \Delta \rho=-0.0006\pm 0.0008,
 \label{eq:Delrho_exp}
\end{equation}
which cuts the allowed parameter space for the Ma-model. Since we want to focus on neutrino physics and lepton flavour violation, we do not try to fit the Higgs sector as well, but rather use four different benchmark scenarios that all fulfill the consistency conditions, as well as the experimental bounds from direct searches and from the measurement of the correction to the $\rho$-parameter (at 3$\sigma$). In the form $(m_1,m_2,\lambda_1,\lambda_2,\lambda_3,\lambda_4,\lambda_5)$, these scenarios are:
\begin{eqnarray}
 \alpha: && (100i{\rm GeV},   75{\rm GeV}, 0.24, 0.10, 0.10, -0.15, -0.10)\nonumber\\
 \beta:  && (100i{\rm GeV}, 98.5{\rm GeV}, 0.24, 0.30, 0.09, -0.18, -0.11)\nonumber\\
 \gamma: && (100i{\rm GeV},  950{\rm GeV}, 0.24, 0.50, 0.02, -0.12, -0.10)\nonumber\\
 \delta: && (100i{\rm GeV},  550{\rm GeV}, 0.24, 0.30, 0.02, -0.05, -0.01)
 \label{eq:Higgs-scenarios}
\end{eqnarray}
The corresponding Higgs masses are given in Tab.~\ref{tab:Higgs-masses}. We have chosen these four scenarios such that they are consistent with the 3$\sigma$-range of WMAP-data for $H^0$ being the Dark Matter candidate, which cuts the allowed parameter space significantly~\cite{Dolle:2009fn}. This leads to some more consistency conditions, as $H^0$ has to be the lightest of all scalars and it also has to be lighter than the heavy right-handed neutrinos.

For all these scenarios, we do the following:
\begin{enumerate}

\item First, the models are fitted to neutrino oscillation data, i.e., mixing angles and mass square differences~\cite{GonzalezGarcia:2007ib}. This is done by the $\chi^2$-function
\begin{equation}
 \chi^2=\sum_{i=1}^{N}\frac{(q_i-q_i^{\rm exp})^2}{\sigma_i^2},
 \label{eq:chi2}
\end{equation}
where $q_i$ are the observables obtained from neutrino oscillations ($\theta_{12}$, $\theta_{13}$, $\theta_{23}$, $\Delta m_A^2$, $\Delta m_\odot^2$), which are calculated in terms of the model parameters (cf.\ Sec.~\ref{sec:models}). $q_i^{\rm exp}$ are their measured counterparts and $\sigma_i$ are the corresponding (symmetrized) standard deviations. The best-fit model parameters are determined by a minimization of the $\chi^2$-function. By projection on the different directions in the parameter space, we determine the $1\sigma$- and $3\sigma$-ranges of the model parameters.

\item Next, we calculate the maximum and minimum values of the quantities measured in different LFV-experiments ($\mu\rightarrow e\gamma$, $\tau\rightarrow \mu\gamma$, $\tau\rightarrow e\gamma$, and $\mu$-$e$ conversion for four different nuclei) by varying the model parameters within their 1$\sigma$- and 3$\sigma$-ranges.

\item Finally, we compare how well different past and future LFV-experiments are able to constrain or exclude the particular model in the four scenarios.

\end{enumerate}

\subsubsection{\label{sec:chi2} The $\chi^2$-fit}

After outlining the general points, we will explain the procedure in more detail using scenario $\alpha$ (cf.\ Eq.~\eqref{eq:Higgs-scenarios}) in connection with Model 1 (cf.\ Sec.~\ref{sec:A4}) as example.

The $\chi^2$-function has already been given in Eq.~\eqref{eq:chi2} and the experimental values and errors of the neutrino observables are summarized in Tab.~\ref{tab:neutrino_params}. These observables in terms of model parameters have been given in Eq.~\eqref{eq:neutrino-observables_1}. The minimization of the $\chi^2$-function then yields the following best-fit values for the three parameters:
\begin{equation}
 a=0.0189,\ b=-0.691,\ M=2.42\cdot 10^6~{\rm GeV}.
 \label{eq:model_1_BF}
\end{equation}
Note that the parameter $b$ is negative to fit the normal mass ordering, cf.\ Eq.~\eqref{eq:neutrino-observables_1}. In the minimization we have required $M_{1,2,3}>m(H^0)$ and $M_{1,2,3}>M_Z/2$ for consistency reasons.

The 1$\sigma$-(3$\sigma$-) values for the model parameters are obtained by inserting all values from Eq.~\eqref{eq:model_1_BF} into the $\chi^2$-function, except for the one parameter that is to be constrained, and by determining the intersections of the remaining 1-dimensional function $\Delta \chi^2\equiv\chi^2-\chi^2_{\rm min}$ with 1(9). For the above parameters, this yields in the form $^{+1\sigma,+3\sigma}_{-1\sigma,-3\sigma}$:
\begin{eqnarray}
 a: && ^{+0.0003,+0.0009}_{-0.0003,-0.0009},\nonumber\\
 b: && ^{+0.003,+0.009}_{-0.003,-0.009},\nonumber\\
 M: && ^{+0.02,+0.05}_{-0.02,-0.05}\cdot 10^6~{\rm GeV}.
 \label{eq:model_1_ranges}
\end{eqnarray}
These are the ranges that we will use in the subsequent analysis. Note that in this model, they are already quite narrow, which is a manifestation of the fact that this model holds a lot of structure.

\subsubsection{\label{sec:LFV-values} Predictions for various LFV-experiments}

The most important types of LFV-experiments are rare lepton decays, $e_i\rightarrow e_j \gamma$, as well as conversions of a bound muon to an electron for some nucleus $N$, $\mu N \rightarrow e N$. In a Ma-like model, the decisive quantities for both types of processes are~\cite{Lavoura:2003xp} ($ij=e_i\rightarrow e_j \gamma/\textrm{$e_i$-$e_j$-conversion}$):
\begin{equation}
 \sigma_{ij}\equiv \frac{-i}{2 m^2(H^\pm)} \sum_{k=1}^3 h_{jk}^* h_{ik} \left[ (m_i+m_j) I_a \left( \frac{M_k^2}{m^2(H^\pm)} \right) +M_k I_b \left( \frac{M_k^2}{m^2(H^\pm)} \right)  \right],
 \label{eq:LFV-sigma}
\end{equation}
where
\begin{equation}
 I_a(t)=\frac{1}{16\pi^2} \left[ \frac{2t^2 +5t-1}{12(t-1)^3} -\frac{t^2 \ln t}{2 (t-1)^4} \right]\ {\rm and}\ I_b(t)=\frac{1}{16\pi^2} \left[ \frac{t+1}{2(t-1)^2} -\frac{t \ln t}{(t-1)^3} \right].
 \label{eq:loop-functions_mueg}
\end{equation}
Using these, the branching ratios for the processes are given by
\begin{equation}
 {\rm Br} (e_i \rightarrow e_j \gamma)= \frac{m_i^3}{8\pi} \frac{|\sigma_{ij}|^2}{\Gamma (e_i \rightarrow e_j \nu_i \overline{\nu}_j)}\ {\rm and}\ {\rm Br} (\mu N \rightarrow e N)=\frac{\pi^2}{2^5 m_\mu^2} \frac{D_N^2}{\omega_{\rm capt} (N)} |\sigma_{\mu e}|^2.
 \label{eq:decay-widths}
\end{equation}
In the first formula, we have neglected the final state lepton mass. The quantities $D_N$ and $\omega_{\rm capt} (N)$, as well as a general expression for the second formula are given in Ref.~\cite{Kitano:2002mt}.

\subsubsection{\label{sec:LFV-experiments} Past and future LFV-experiments for Model 1}

We then use the parameter ranges from Eq.~\eqref{eq:model_1_ranges} to make predictions with Eq.~\eqref{eq:decay-widths}. The result is included in Fig.~\ref{fig:model_1}. Furthermore, we have put in the limits/sensitivities of several past/future experiments, all listed in Tab.~\ref{tab:LFV-experiments}. A further discussion of the results will be given in the next section.

\section{\label{sec:results} Results}

\begin{table}[t]
\centering
\begin{tabular}[h]{|c||c|c|c|c|}\hline
Experiment  & Status    & Process                            & BR-Limit/Sensitivity\\ \hline \hline
MEGA        & Past      & $\mu\rightarrow e\gamma$           &  $1.2\cdot 10^{-11}$ \\ \hline
MEG         & Future    & $\mu\rightarrow e\gamma$           &  $1.0\cdot 10^{-13}$ \\ \hline
BELLE       & Past      & $\tau\rightarrow \mu\gamma$        &  $4.5\cdot 10^{-8}$ \\ \hline
Babar       & Past      & $\tau\rightarrow e\gamma$          &  $1.1\cdot 10^{-7}$ \\ \hline
MECO        & Cancelled & $\mu{\rm Al}\rightarrow e{\rm Al}$ &  $2.0\cdot 10^{-17}$ \\ \hline
SINDRUM II  & Past      & $\mu{\rm Ti}\rightarrow e{\rm Ti}$ &  $6.1\cdot 10^{-13}$ \\ \hline
PRISM/PRIME & Future    & $\mu{\rm Ti}\rightarrow e{\rm Ti}$ &  $5.0\cdot 10^{-19}$ \\ \hline
SINDRUM II  & Past      & $\mu{\rm Au}\rightarrow e{\rm Au}$ &  $7.0\cdot 10^{-13}$ \\ \hline
SINDRUM II  & Past      & $\mu{\rm Pb}\rightarrow e{\rm Pb}$ &  $4.6\cdot 10^{-11}$ \\ \hline
\end{tabular}
\caption{\label{tab:LFV-experiments} Limits on the branching ratios for several past and future LFV-experiments~\cite{Raidal:2008jk}.}
\end{table}

\begin{figure}[t]
\centering
\begin{tabular}[h]{lr}
\epsfig{file=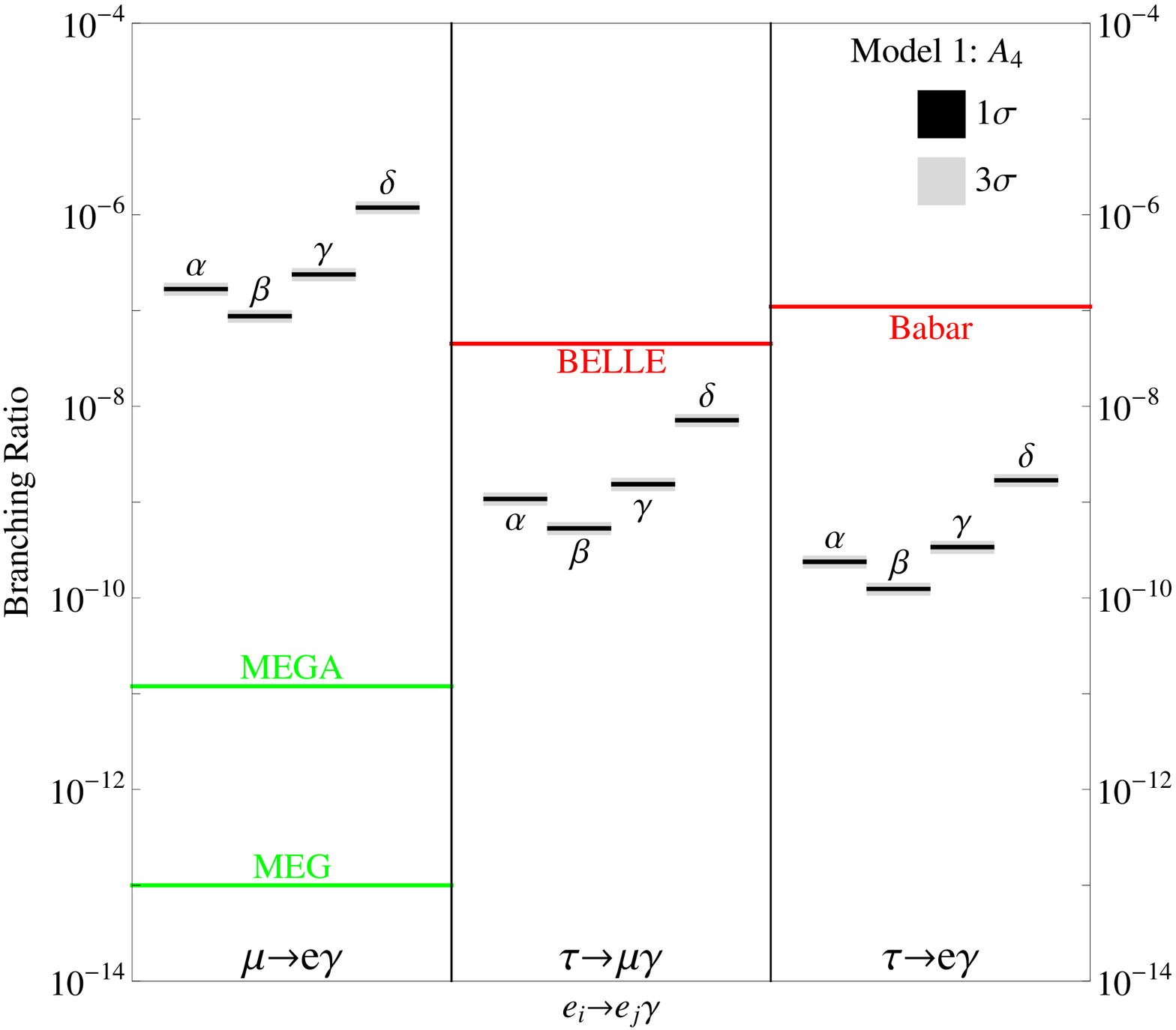,width=8cm}  &
\epsfig{file=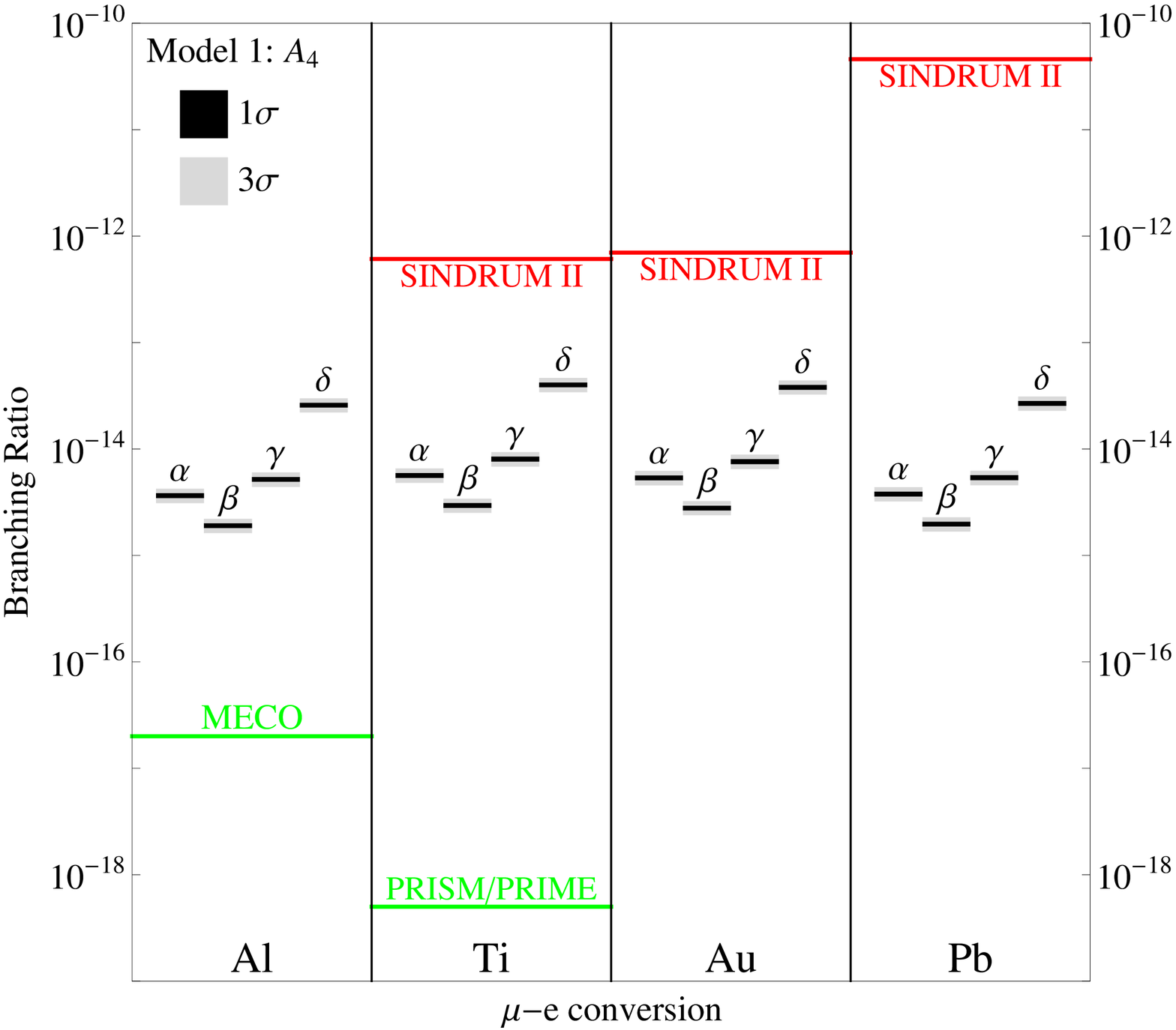,width=8cm}
\end{tabular}
\caption{\label{fig:model_1} The numerical results of our analysis for Model 1.}
\end{figure}

\begin{figure}[t]
\centering
\begin{tabular}[h]{lr}
\epsfig{file=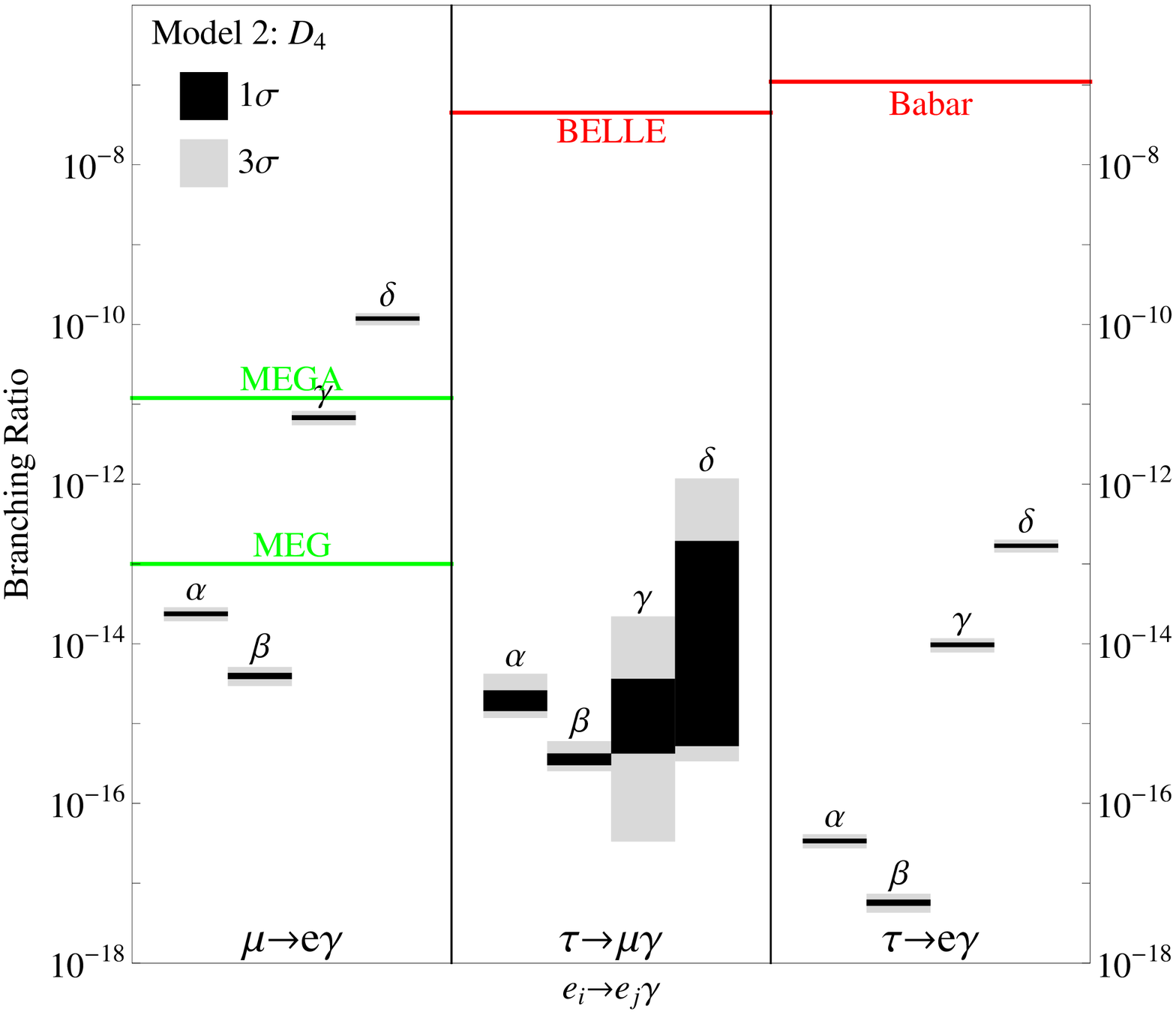,width=8cm}  &
\epsfig{file=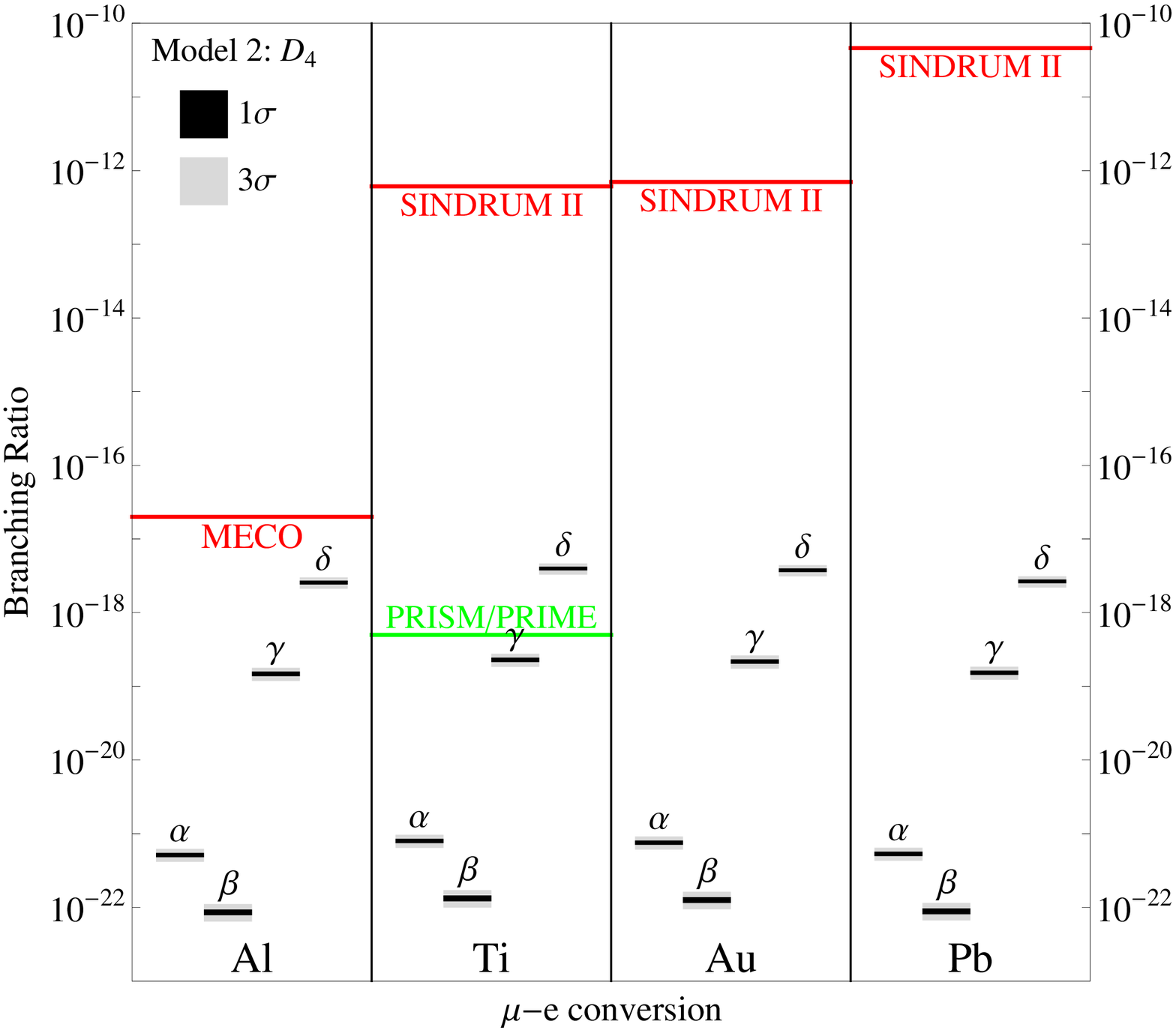,width=8cm}
\end{tabular}
\caption{\label{fig:model_2} The numerical results of our analysis for Model 2.}
\end{figure}

We will now discuss how the general conflict between an extended scalar sector and flavour symmetries looks in our example models. Let us first start with Model 1. The numerical results can be seen in Fig.~\ref{fig:model_1}: On the left panel, we present the $1\sigma$ (black) and $3\sigma$ (gray) predictions of Model 1 for the processes $\mu \rightarrow e\gamma$, $\tau \rightarrow \mu\gamma$, and $\tau \rightarrow e\gamma$, as well as different present and future bounds from several experiments, cf.\ Tab.~\ref{tab:LFV-experiments}. The right panel shows the same for $\mu$-$e$ conversion on the elements Al, Ti, Au, and Pb.

Model 1 is the prime example that our logic works: As explained in Sec.~\ref{sec:A4}, there are only 3 free parameters in the model. Still, it is able to fit the neutrino data well. Actually, the only deviations from a perfect fit arise from the very accurate prediction of the mixing angles (e.g., the experimental best-fit value of $\theta_{23}$ is not exactly maximal; cf.\ Eq.~\eqref{eq:neutrino-observables_1} and Tab.~\ref{tab:neutrino_params}). The obtained parameter ranges are, however, quite narrow, as can be seen from the example given in Sec.~\ref{sec:chi2}. This is exactly the point, where the experimental limits on LFV-processes get really powerful: Because of the stiffness in the model parameter space, the prediction of, e.g., the branching ratio $\mu \rightarrow e\gamma$ is so clear, that only a very narrow window is left for parameter variations. Accordingly, this model is actually already excluded by the past MEGA experiment (cf.\ Fig.~\ref{fig:model_1}) for all four Higgs scenarios from Eq.~\eqref{eq:Higgs-scenarios}. We want to stress again, that these four scenarios belong to the few regions in parameter space that are indeed consistent with all the data and constraints mentioned in Sec.~\ref{sec:general}. The branching ratios for $\mu$-$e$ conversion are in general lower, and pass all current constraints. However, in this sector PRISM/PRIME will provide another future bound that will be able to exclude this model.

The remaining questions is how far we can stretch this logic for models with less and less predictivity. As example for that case we can use Model 2, which has seven free parameters to fit the data (cf.\ Sec.~\ref{sec:D4}). This more than doubles the degrees of freedom in the fit.

The numerical results for this model are given in Fig.~\ref{fig:model_2}. First of all, it may look odd that here, all $1\sigma$ and $3\sigma$ regions are somehow narrow, except for $\tau \rightarrow \mu \gamma$. This is simply because all branching ratios are essentially functions of the product $|ab|$ (where $a$ and $b$ are model parameters), while the one for $\tau \rightarrow \mu \gamma$ is given by the sum of three contributions, which are proportional to $|b|^2$, $|c|^2$, and $|d|^2$, respectively. This numerical example nicely shows how more freedom blows up the regions which are predicted by a certain model. Turning this argumentation round, a certain limit on some observable is weaker the more free parameters there are that influence the observable in question.

However, even this model with much less predictivity than the one before can be excluded for some scenarios: Scenario~$\delta$ has already been excluded by the MEGA-experiment and scenario~$\gamma$ can be tested by MEG. This shows the strength of our considerations: Even for a model that has a lot of freedom our logic still applies in suitable settings, which are here given by the scenarios~$\gamma$ and~$\delta$. Actually, even the scenarios~$\alpha$ and~$\beta$ are not that far below the future MEG-bound, and especially a hypothetical future experiment aiming at $\tau \rightarrow \mu \gamma$ might be very suitable to exclude this particular model.

\section{\label{sec:conclusions} Conclusions}

In this paper we have studied the conflict arising in models with an extended scalar sector and discrete flavour symmetries when confronted with LFV-bounds. We have illustrated this using two examples based on the Ma-model, one with an $A_4$ and with a $D_4$ symmetry. Since the first model exhibits a relatively rigid structure (only three free parameters), it is already excluded for all four scenarios by existing bounds. Even though the second model has more than twice as many free parameters, it can still be strongly constrained and two of the scenarios can either be excluded or tested in the near future.

We want to stress, however, that our considerations are not at all restricted to Ma-like models, but should apply to a much wider class of theories. Models with a lot of structure (meaning few parameters) may easily be excluded by existing or future LFV-bounds although they have no problems without the flavour symmetry. Even models with many parameters can at least be strongly constrained, if not excluded as well.

\section*{\label{sec:Ack}Acknowledgments} 

We would like to thank W.~Rodejohann and especially A.~Blum for useful discussions. This work has been supported by the DFG-Sonderforschungsbereich Transregio 27 ``Neutrinos and beyond -- Weakly interacting particles in Physics, Astrophysics and Cosmology''.

\newpage
\section*{\label{GroupA4}Appendix A: Group Theory of $A_4$ \cite{GAFF} }

\renewcommand{\theequation}{A-\arabic{equation}}
\setcounter{equation}{0}  

The group $A_4$ is a group which describes even permutations of four objects. It has two generators, $S$ and $T$, that fulfill the relations
\begin{equation}
S^2=(ST)^3=T^3=1.
\end{equation}
The group has four inequivalent irreducible representations, $\Rep{1},\Rep{1}',\Rep{1}''$, and $\Rep{3}$, which transform under the generators, $S$ and $T$ as follows:
\begin{eqnarray}
\Rep{1}: & & S=1, \;\; T=1, \nonumber \\
\Rep{1}': & & S=1, \;\; T=\omega^2, \nonumber \\
\Rep{1}'': & & S=1, \;\; T=\omega,
\end{eqnarray}
\begin{equation}
\Rep{3}: \;\;\; T=\begin{pmatrix}
 1 & 0 & 0\\
 0 & \omega^2 & 0\\
 0 & 0 & \omega
 \end{pmatrix}, \;\;\; 
S=\frac{1}{3}\begin{pmatrix}
 -1 & 2 & 2\\
 2 & -1 & 2\\
 2 & 2 & -1
 \end{pmatrix},
\end{equation}
where $\omega=e^{i 2\pi/3}$ (which implies $\omega^4=\omega$). \\
The product rules for the singlets are the following:
\begin{equation}
\Rep{1}' \times \Rep{1}'= \Rep{1}'' \;, \;\;\; \Rep{1}' \times \Rep{1}''= \Rep{1} \;, \;\;\; \Rep{1}'' \times \Rep{1}''= \Rep{1}' \;, \;\;\; \Rep{1} \times \Rep{1}= \Rep{1} \;, \;\;\; \Rep{1} \times \Rep{1}'= \Rep{1}'\;, \;\;\; \Rep{1} \times \Rep{1}''= \Rep{1}''.
\end{equation} 
Consider now two triplets:
\begin{equation}
a=(a_1,a_2,a_3)^T \;, \;\;\;\; b=(b_1,b_2,b_3)^T.
\end{equation}
The product of these two triplets can be decomposed as
\begin{equation}
\Rep{3} \times \Rep{3} = \Rep{1}+\Rep{1}'+\Rep{1}''+\Rep{3}_s+\Rep{3}_a,
\end{equation}
where
\begin{eqnarray}
\Rep{1} &=& (ab) =a_1b_1+a_2b_3+a_3b_2, \nonumber \\
\Rep{1}' &=& (ab)' = a_3b_3+a_1b_2+a_2b_1, \nonumber \\
\Rep{1}'' &=& (ab)'' = a_2b_2+a_1b_3+a_3b_1, \nonumber \\
\end{eqnarray}
and
\begin{eqnarray}
\Rep{3}_s &=& (ab)_s =\frac{1}{2}(2a_1b_1-a_2b_3-a_3b_2,2a_3b_3-a_1b_2-a_2b_1,2a_2b_2-a_1b_3-a_3b_1)^T, \nonumber \\
\Rep{3}_a &=& (ab)_a = \frac{1}{2}(a_2b_3-a_3b_2,a_1b_2-a_2b_1,a_1b_3-a_3b_1)^T.
\end{eqnarray}

\section*{\label{GroupD4}Appendix B: Group Theory of $D_4$ \cite{D4paper, Dnpaper}}

\renewcommand{\theequation}{B-\arabic{equation}}
\setcounter{equation}{0}  

The group $D_4$ is a group which describes the symmetry of a square. It has two generators, $A$ and $B$, that fulfill the relations
\begin{equation}
A^4=B^2=1 \;\;\;\mbox{and} \;\;\; ABA=B.
\end{equation}
The irreducible representations consist of four singlets, $\MoreRep{1}{1},\MoreRep{1}{2},\MoreRep{1}{3},\MoreRep{1}{4}$, and one doublet $\Rep{2}$, which transform under the generators, $A$ and $B$ as follows:
\begin{eqnarray}
\MoreRep{1}{1}: & & A=1, \;\; B=1, \nonumber \\
\MoreRep{1}{2}: & & A=1, \;\; B=-1, \nonumber \\
\MoreRep{1}{3}: & & A=-1, \;\; B=1, \nonumber \\
\MoreRep{1}{4}: & & A=-1, \;\; B=-1,
\end{eqnarray}
\begin{equation}
\Rep{2}: \;\;\; A=\begin{pmatrix}
 i & 0 \\
 0 & -i
\end{pmatrix}, \;\;\; 
B=\begin{pmatrix}
 0 & 1 \\
 1 & 0
 \end{pmatrix}.
\end{equation}
The product rules for $\MoreRep{1}{i}$ are
\begin{eqnarray}
&&
\MoreRep{1}{i} \times \MoreRep{1}{i}= \MoreRep{1}{1} \; , \;\; 
\MoreRep{1}{1} \times \MoreRep{1}{i}= \MoreRep{1}{i} \;\; \mbox{for} \;\; \rm i=1,...,4
\; , \;\;
\MoreRep{1}{2} \times \MoreRep{1}{3}= \MoreRep{1}{4} \; , \;\;
\MoreRep{1}{2} \times \MoreRep{1}{4}= \MoreRep{1}{3} \; , \;\; \mbox{and}\nonumber\\
&&
\MoreRep{1}{3} \times \MoreRep{1}{4}= \MoreRep{1}{2} \; .
\end{eqnarray}
For $s_i \sim \MoreRep{1}{i}$ and $(a_1,a_2)^{T} \sim \Rep{2}$ we find
\begin{equation}
\left( \begin{array}{c} s_1 a_1 \\ s_1 a_2
\end{array} \right) \sim \Rep{2} \;\; , \;\;\;
\left( \begin{array}{c} s_2 a_1 \\ -s_2 a_2
\end{array} \right) \sim \Rep{2} \;\; , \;\;\;
\left( \begin{array}{c} s_3 a_2 \\ s_3 a_1
\end{array} \right) \sim \Rep{2}  \; , \;\; \mbox{and} \;\;\;
\left( \begin{array}{c} s_4 a_2 \\ -s_4 a_1
\end{array} \right) \sim \Rep{2} \;\; .
\end{equation}
For $(a_1,a_2)^{T}$, $(b_1, b_2)^{T}$ $\sim \Rep{2}$, the product $\Rep{2} \times \Rep{2}$ decomposes into the four singlets which read 
\begin{equation}
a_1 b_2 + a_2 b_1 \sim \MoreRep{1}{1} \;\; , \;\;\;
a_1 b_2 - a_2 b_1 \sim \MoreRep{1}{2} \;\; , \;\;\;
a_1 b_1 + a_2 b_2 \sim \MoreRep{1}{3} \;\;\; \mbox{and} \;\;\;
a_1 b_1 - a_2 b_2 \sim \MoreRep{1}{4} \;\; .
\end{equation}

\newpage
\bibliographystyle{./apsrev}
\bibliography{./DiscreteLFV}

\begin{thebibliography}{10}
\expandafter\ifx\csname bibnamefont\endcsname\relax
  \def\bibnamefont#1{#1}\fi
\expandafter\ifx\csname bibfnamefont\endcsname\relax
  \def\bibfnamefont#1{#1}\fi
\expandafter\ifx\csname url\endcsname\relax
  \def\url#1{\texttt{#1}}\fi
\expandafter\ifx\csname urlprefix\endcsname\relax\def\urlprefix{URL }\fi
\providecommand{\bibinfo}[2]{#2}
\providecommand{\eprint}[2][]{\url{#2}}

\bibitem{Bertone:2004pz}
\bibinfo{author}{\bibfnamefont{G.}~\bibnamefont{Bertone}},
  \bibinfo{author}{\bibfnamefont{D.}~\bibnamefont{Hooper}}, \bibnamefont{and}
  \bibinfo{author}{\bibfnamefont{J.}~\bibnamefont{Silk}},
  \bibinfo{journal}{Phys. Rept.} \textbf{\bibinfo{volume}{405}},
  \bibinfo{pages}{279} (\bibinfo{year}{2005}), \eprint{hep-ph/0404175}.

\bibitem{Riotto:1998bt}
\bibinfo{author}{\bibfnamefont{A.}~\bibnamefont{Riotto}}
  (\bibinfo{year}{1998}), \eprint{hep-ph/9807454}.

\bibitem{Martin:1997ns}
\bibinfo{author}{\bibfnamefont{S.~P.} \bibnamefont{Martin}}
  (\bibinfo{year}{1997}), \eprint{hep-ph/9709356}.

\bibitem{Barger:2007im}
\bibinfo{author}{\bibfnamefont{V.}~\bibnamefont{Barger}},
  \bibinfo{author}{\bibfnamefont{P.}~\bibnamefont{Langacker}},
  \bibinfo{author}{\bibfnamefont{M.}~\bibnamefont{McCaskey}},
  \bibinfo{author}{\bibfnamefont{M.~J.} \bibnamefont{Ramsey-Musolf}},
  \bibnamefont{and}
  \bibinfo{author}{\bibfnamefont{G.}~\bibnamefont{Shaughnessy}},
  \bibinfo{journal}{Phys. Rev.} \textbf{\bibinfo{volume}{D77}},
  \bibinfo{pages}{035005} (\bibinfo{year}{2008}), \eprint{0706.4311}.

\bibitem{Gunion:1989we}
\bibinfo{author}{\bibfnamefont{J.~F.} \bibnamefont{Gunion}},
  \bibinfo{author}{\bibfnamefont{H.~E.} \bibnamefont{Haber}},
  \bibinfo{author}{\bibfnamefont{G.~L.} \bibnamefont{Kane}}, \bibnamefont{and}
  \bibinfo{author}{\bibfnamefont{S.}~\bibnamefont{Dawson}}
  \bibinfo{note}{SCIPP-89/13}.

\bibitem{Haber:2006ue}
\bibinfo{author}{\bibfnamefont{H.~E.} \bibnamefont{Haber}} \bibnamefont{and}
  \bibinfo{author}{\bibfnamefont{D.}~\bibnamefont{O'Neil}},
  \bibinfo{journal}{Phys. Rev.} \textbf{\bibinfo{volume}{D74}},
  \bibinfo{pages}{015018} (\bibinfo{year}{2006}), \eprint{hep-ph/0602242}.

\bibitem{Diaz:2000cm}
\bibinfo{author}{\bibfnamefont{R.}~\bibnamefont{Diaz}},
  \bibinfo{author}{\bibfnamefont{R.}~\bibnamefont{Martinez}}, \bibnamefont{and}
  \bibinfo{author}{\bibfnamefont{J.~A.} \bibnamefont{Rodriguez}},
  \bibinfo{journal}{Phys. Rev.} \textbf{\bibinfo{volume}{D63}},
  \bibinfo{pages}{095007} (\bibinfo{year}{2001}), \eprint{hep-ph/0010149}.

\bibitem{Blum:2007he}
\bibinfo{author}{\bibfnamefont{A.}~\bibnamefont{Blum}} \bibnamefont{and}
  \bibinfo{author}{\bibfnamefont{A.}~\bibnamefont{Merle}},
  \bibinfo{journal}{Phys. Rev.} \textbf{\bibinfo{volume}{D77}},
  \bibinfo{pages}{076005} (\bibinfo{year}{2008}), \eprint{0709.3294}.

\bibitem{Raidal:2008jk}
\bibinfo{author}{\bibfnamefont{M.}~\bibnamefont{Raidal}} \emph{et~al.},
  \bibinfo{journal}{Eur. Phys. J.} \textbf{\bibinfo{volume}{C57}},
  \bibinfo{pages}{13} (\bibinfo{year}{2008}), \eprint{0801.1826}.

\bibitem{Grimus:2005mu}
\bibinfo{author}{\bibfnamefont{W.}~\bibnamefont{Grimus}} \bibnamefont{and}
  \bibinfo{author}{\bibfnamefont{L.}~\bibnamefont{Lavoura}},
  \bibinfo{journal}{JHEP} \textbf{\bibinfo{volume}{08}}, \bibinfo{pages}{013}
  (\bibinfo{year}{2005}), \eprint{hep-ph/0504153}.

\bibitem{Grimus:2003kq}
\bibinfo{author}{\bibfnamefont{W.}~\bibnamefont{Grimus}} \bibnamefont{and}
  \bibinfo{author}{\bibfnamefont{L.}~\bibnamefont{Lavoura}},
  \bibinfo{journal}{Phys. Lett.} \textbf{\bibinfo{volume}{B572}},
  \bibinfo{pages}{189} (\bibinfo{year}{2003}), \eprint{hep-ph/0305046}.

\bibitem{Ishimori:2008gp}
\bibinfo{author}{\bibfnamefont{H.}~\bibnamefont{Ishimori}} \emph{et~al.},
  \bibinfo{journal}{Phys. Lett.} \textbf{\bibinfo{volume}{B662}},
  \bibinfo{pages}{178} (\bibinfo{year}{2008}), \eprint{0802.2310}.

\bibitem{D4paper}
\bibinfo{author}{\bibfnamefont{A.}~\bibnamefont{Adulpravitchai}},
  \bibinfo{author}{\bibfnamefont{A.}~\bibnamefont{Blum}}, \bibnamefont{and}
  \bibinfo{author}{\bibfnamefont{C.}~\bibnamefont{Hagedorn}},
  \bibinfo{journal}{JHEP} \textbf{\bibinfo{volume}{03}}, \bibinfo{pages}{046}
  (\bibinfo{year}{2009}), \eprint{0812.3799}.

\bibitem{Fukuyama:1997ky}
\bibinfo{author}{\bibfnamefont{T.}~\bibnamefont{Fukuyama}} \bibnamefont{and}
  \bibinfo{author}{\bibfnamefont{H.}~\bibnamefont{Nishiura}}
  (\bibinfo{year}{1997}), \eprint{hep-ph/9702253}.

\bibitem{Ma:2001dn}
\bibinfo{author}{\bibfnamefont{E.}~\bibnamefont{Ma}} \bibnamefont{and}
  \bibinfo{author}{\bibfnamefont{G.}~\bibnamefont{Rajasekaran}},
  \bibinfo{journal}{Phys. Rev.} \textbf{\bibinfo{volume}{D64}},
  \bibinfo{pages}{113012} (\bibinfo{year}{2001}), \eprint{hep-ph/0106291}.

\bibitem{Babu:2002dz}
\bibinfo{author}{\bibfnamefont{K.~S.} \bibnamefont{Babu}},
  \bibinfo{author}{\bibfnamefont{E.}~\bibnamefont{Ma}}, \bibnamefont{and}
  \bibinfo{author}{\bibfnamefont{J.~W.~F.} \bibnamefont{Valle}},
  \bibinfo{journal}{Phys. Lett.} \textbf{\bibinfo{volume}{B552}},
  \bibinfo{pages}{207} (\bibinfo{year}{2003}), \eprint{hep-ph/0206292}.

\bibitem{Altarelli:2005yp}
\bibinfo{author}{\bibfnamefont{G.}~\bibnamefont{Altarelli}} \bibnamefont{and}
  \bibinfo{author}{\bibfnamefont{F.}~\bibnamefont{Feruglio}},
  \bibinfo{journal}{Nucl. Phys.} \textbf{\bibinfo{volume}{B720}},
  \bibinfo{pages}{64} (\bibinfo{year}{2005}), \eprint{hep-ph/0504165}.

\bibitem{GAFF}
\bibinfo{author}{\bibfnamefont{G.}~\bibnamefont{Altarelli}} \bibnamefont{and}
  \bibinfo{author}{\bibfnamefont{F.}~\bibnamefont{Feruglio}},
  \bibinfo{journal}{Nucl.Phys} \textbf{\bibinfo{volume}{B741}},
  \bibinfo{pages}{215} (\bibinfo{year}{2006}), \eprint{hep-ph/0512103}.

\bibitem{Altarelli:2006kg}
\bibinfo{author}{\bibfnamefont{G.}~\bibnamefont{Altarelli}},
  \bibinfo{author}{\bibfnamefont{F.}~\bibnamefont{Feruglio}}, \bibnamefont{and}
  \bibinfo{author}{\bibfnamefont{Y.}~\bibnamefont{Lin}},
  \bibinfo{journal}{Nucl. Phys.} \textbf{\bibinfo{volume}{B775}},
  \bibinfo{pages}{31} (\bibinfo{year}{2007}), \eprint{hep-ph/0610165}.

\bibitem{Harrison:2002er}
\bibinfo{author}{\bibfnamefont{P.~F.} \bibnamefont{Harrison}},
  \bibinfo{author}{\bibfnamefont{D.~H.} \bibnamefont{Perkins}},
  \bibnamefont{and} \bibinfo{author}{\bibfnamefont{W.~G.} \bibnamefont{Scott}},
  \bibinfo{journal}{Phys. Lett.} \textbf{\bibinfo{volume}{B530}},
  \bibinfo{pages}{167} (\bibinfo{year}{2002}), \eprint{hep-ph/0202074}.

\bibitem{Ma:2006km}
\bibinfo{author}{\bibfnamefont{E.}~\bibnamefont{Ma}}, \bibinfo{journal}{Phys.
  Rev.} \textbf{\bibinfo{volume}{D73}}, \bibinfo{pages}{077301}
  (\bibinfo{year}{2006}), \eprint{hep-ph/0601225}.

\bibitem{Sierra:2008wj}
\bibinfo{author}{\bibfnamefont{D.}~\bibnamefont{Aristizabal~Sierra}},
  \bibinfo{author}{\bibfnamefont{J.}~\bibnamefont{Kubo}},
  \bibinfo{author}{\bibfnamefont{D.}~\bibnamefont{Restrepo}},
  \bibinfo{author}{\bibfnamefont{D.}~\bibnamefont{Suematsu}}, \bibnamefont{and}
  \bibinfo{author}{\bibfnamefont{O.}~\bibnamefont{Zapata}},
  \bibinfo{journal}{Phys. Rev.} \textbf{\bibinfo{volume}{D79}},
  \bibinfo{pages}{013011} (\bibinfo{year}{2009}), \eprint{0808.3340}.

\bibitem{Lindner:2005kr}
\bibinfo{author}{\bibfnamefont{M.}~\bibnamefont{Lindner}},
  \bibinfo{author}{\bibfnamefont{A.}~\bibnamefont{Merle}}, \bibnamefont{and}
  \bibinfo{author}{\bibfnamefont{W.}~\bibnamefont{Rodejohann}},
  \bibinfo{journal}{Phys. Rev.} \textbf{\bibinfo{volume}{D73}},
  \bibinfo{pages}{053005} (\bibinfo{year}{2006}), \eprint{hep-ph/0512143}.

\bibitem{Ma:2009dk}
\bibinfo{author}{\bibfnamefont{E.}~\bibnamefont{Ma}}  (\bibinfo{year}{2009}),
  \eprint{0905.0221}.

\bibitem{Zee:1980ai}
\bibinfo{author}{\bibfnamefont{A.}~\bibnamefont{Zee}}, \bibinfo{journal}{Phys.
  Lett.} \textbf{\bibinfo{volume}{B93}}, \bibinfo{pages}{389}
  (\bibinfo{year}{1980}).

\bibitem{Babu:1988ig}
\bibinfo{author}{\bibfnamefont{K.~S.} \bibnamefont{Babu}} \bibnamefont{and}
  \bibinfo{author}{\bibfnamefont{E.}~\bibnamefont{Ma}}, \bibinfo{journal}{Phys.
  Rev. Lett.} \textbf{\bibinfo{volume}{61}}, \bibinfo{pages}{674}
  (\bibinfo{year}{1988}).

\bibitem{Zee:1985id}
\bibinfo{author}{\bibfnamefont{A.}~\bibnamefont{Zee}}, \bibinfo{journal}{Nucl.
  Phys.} \textbf{\bibinfo{volume}{B264}}, \bibinfo{pages}{99}
  (\bibinfo{year}{1986}).

\bibitem{Babu:1988ki}
\bibinfo{author}{\bibfnamefont{K.~S.} \bibnamefont{Babu}},
  \bibinfo{journal}{Phys. Lett.} \textbf{\bibinfo{volume}{B203}},
  \bibinfo{pages}{132} (\bibinfo{year}{1988}).

\bibitem{Suematsu:2009ww}
\bibinfo{author}{\bibfnamefont{D.}~\bibnamefont{Suematsu}},
  \bibinfo{author}{\bibfnamefont{T.}~\bibnamefont{Toma}}, \bibnamefont{and}
  \bibinfo{author}{\bibfnamefont{T.}~\bibnamefont{Yoshida}}
  (\bibinfo{year}{2009}), \eprint{0903.0287}.

\bibitem{Dolle:2009fn}
\bibinfo{author}{\bibfnamefont{E.~M.} \bibnamefont{Dolle}} \bibnamefont{and}
  \bibinfo{author}{\bibfnamefont{S.}~\bibnamefont{Su}}, \bibinfo{journal}{Phys.
  Rev.} \textbf{\bibinfo{volume}{D80}}, \bibinfo{pages}{055012}
  (\bibinfo{year}{2009}), \eprint{0906.1609}.

\bibitem{Eriksson:2009ws}
\bibinfo{author}{\bibfnamefont{D.}~\bibnamefont{Eriksson}},
  \bibinfo{author}{\bibfnamefont{J.}~\bibnamefont{Rathsman}}, \bibnamefont{and}
  \bibinfo{author}{\bibfnamefont{O.}~\bibnamefont{Stal}}
  (\bibinfo{year}{2009}), \eprint{0902.0851}.

\bibitem{A4seesawMuandKing}
\bibinfo{author}{\bibfnamefont{M.~C.} \bibnamefont{Chen}} \bibnamefont{and}
  \bibinfo{author}{\bibfnamefont{S.~F.} \bibnamefont{King}},
  \bibinfo{journal}{JHEP} \textbf{\bibinfo{volume}{06}}, \bibinfo{pages}{072}
  (\bibinfo{year}{2009}), \eprint{0903.0125}.

\bibitem{Hirsch:2008rp}
\bibinfo{author}{\bibfnamefont{M.}~\bibnamefont{Hirsch}},
  \bibinfo{author}{\bibfnamefont{S.}~\bibnamefont{Morisi}}, \bibnamefont{and}
  \bibinfo{author}{\bibfnamefont{J.~W.~F.} \bibnamefont{Valle}},
  \bibinfo{journal}{Phys. Rev.} \textbf{\bibinfo{volume}{D78}},
  \bibinfo{pages}{093007} (\bibinfo{year}{2008}), \eprint{0804.1521}.

\bibitem{GonzalezGarcia:2007ib}
\bibinfo{author}{\bibfnamefont{M.~C.} \bibnamefont{Gonzalez-Garcia}}
  \bibnamefont{and} \bibinfo{author}{\bibfnamefont{M.}~\bibnamefont{Maltoni}},
  \bibinfo{journal}{Phys. Rept.} \textbf{\bibinfo{volume}{460}},
  \bibinfo{pages}{1} (\bibinfo{year}{2008}), \eprint{0704.1800}.

\bibitem{Raspereza:2002ni}
\bibinfo{author}{\bibfnamefont{A.}~\bibnamefont{Raspereza}}
  (\bibinfo{year}{2002}), \eprint{hep-ex/0209021}.

\bibitem{Grimus:2007if}
\bibinfo{author}{\bibfnamefont{W.}~\bibnamefont{Grimus}},
  \bibinfo{author}{\bibfnamefont{L.}~\bibnamefont{Lavoura}},
  \bibinfo{author}{\bibfnamefont{O.~M.} \bibnamefont{Ogreid}},
  \bibnamefont{and} \bibinfo{author}{\bibfnamefont{P.}~\bibnamefont{Osland}},
  \bibinfo{journal}{J. Phys.} \textbf{\bibinfo{volume}{G35}},
  \bibinfo{pages}{075001} (\bibinfo{year}{2008}), \eprint{0711.4022}.

\bibitem{Amsler:2008zzb}
\bibinfo{author}{\bibfnamefont{C.}~\bibnamefont{Amsler}} \emph{et~al.}
  (\bibinfo{collaboration}{Particle Data Group}), \bibinfo{journal}{Phys.
  Lett.} \textbf{\bibinfo{volume}{B667}}, \bibinfo{pages}{1}
  (\bibinfo{year}{2008}).

\bibitem{Lavoura:2003xp}
\bibinfo{author}{\bibfnamefont{L.}~\bibnamefont{Lavoura}},
  \bibinfo{journal}{Eur. Phys. J.} \textbf{\bibinfo{volume}{C29}},
  \bibinfo{pages}{191} (\bibinfo{year}{2003}), \eprint{hep-ph/0302221}.

\bibitem{Kitano:2002mt}
\bibinfo{author}{\bibfnamefont{R.}~\bibnamefont{Kitano}},
  \bibinfo{author}{\bibfnamefont{M.}~\bibnamefont{Koike}}, \bibnamefont{and}
  \bibinfo{author}{\bibfnamefont{Y.}~\bibnamefont{Okada}},
  \bibinfo{journal}{Phys. Rev.} \textbf{\bibinfo{volume}{D66}},
  \bibinfo{pages}{096002} (\bibinfo{year}{2002}), \eprint{hep-ph/0203110}.

\bibitem{Dnpaper}
\bibinfo{author}{\bibfnamefont{A.}~\bibnamefont{Blum}},
  \bibinfo{author}{\bibfnamefont{C.}~\bibnamefont{Hagedorn}}, \bibnamefont{and}
  \bibinfo{author}{\bibfnamefont{M.}~\bibnamefont{Lindner}},
  \bibinfo{journal}{Phys. Rev.} \textbf{\bibinfo{volume}{D77}},
  \bibinfo{pages}{076004} (\bibinfo{year}{2008}), \eprint{0709.3450}.

\end{thebibliography}

\end{document}